# RIOT: I/O-Efficient Numerical Computing without SQL*

Yi Zhang
Duke University
yizhang@cs.duke.edu

Herodotos Herodotou
Duke University
hero@cs.duke.edu

Jun Yang
Duke University
junyang@cs.duke.edu

## ABSTRACT

R is a numerical computing environment that is widely popular for statistical data analysis. Like many such environments, R performs poorly for large datasets whose sizes exceed that of physical memory. We present our vision of *RIOT* (*R with I/O Transparency*), a system that makes R programs I/O-efficient in a way transparent to the users. We describe our experience with *RIOT-DB*, an initial prototype that uses a relational database system as a backend. Despite the overhead and inadequacy of generic database systems in handling array data and numerical computation, RIOT-DB significantly outperforms R in many large-data scenarios, thanks to a suite of high-level, inter-operation optimizations that integrate seamlessly into R. While many techniques in RIOT are inspired by databases (and, for RIOT-DB, realized by a database system), RIOT users are insulated from anything database related. Compared with previous approaches that require users to learn new languages and rewrite their programs to interface with a database, RIOT will, we believe, be easier to adopt by the majority of the R users.

## 1 Introduction

Scientists and engineers rely heavily on numerical computing environments such as R (`http://www.r-project.org/`) and MATLAB. R is an open-source implementation of the S programming language, the de facto standard among statisticians for developing data analysis tools. R has a huge user and developer base; CRAN, the Comprehensive R Archive Network, lists more than 1,500 packages (which extend R's functionalities) available as of September 2008. Like MATLAB, R provides a high level of abstraction to simplify programming of numerical and statistical computation. Most R users learn to program at the level of vectors and matrices instead of using explicit loops to iterate through arrays.

The ever-growing sizes of datasets, however, pose serious challenges to these numerical computing environments. R, for example, assumes that all data fits in main memory. When the physical memory can no longer hold all data, the operating system's virtual memory mechanism starts to swap data to and from disk, often causing the program to thrash and run unbearably slow. This problem stems from the difficulty for the operating system in predicting the program's data access pattern and optimizing I/O accordingly.

When thrashing happens, many R users' first reaction is to rewrite the program to manage I/O explicitly, often in a lower-level language like C or FORTRAN. Rewriting code and hand-optimizing I/O require a huge amount of effort and expertise. Also, what is hand-optimized for one computer may perform poorly on another.

How can we make R programs I/O-efficient without placing so much burden on the programmers? One approach is to use an I/O-efficient library of routines. For example, one could simply replace the implementation of matrix multiply with a library routine that knows how to swap matrix elements in and out efficiently. There are a number of such I/O-efficient numerical computing libraries (e.g., SOLAR [18] and DRA [13]).[1] While developing such libraries is an important first step, we argue that simply having I/O-efficient implementations of individual operations is not enough. Many sources of I/O-inefficiency in a program remain at a higher, inter-operation level: e.g., how intermediate results are passed between operations, how much computation and I/O could be reduced if certain operations are deferred and reordered, etc. We will show in this paper the importance of optimizing I/O at this level.

Database systems seem to be a natural choice to address this problem, because they are I/O-efficient and support SQL, a high-level language that enables advanced optimization. R has packages that provide database connectivity through APIs such as ODBC, but users must be SQL experts to use it effectively for any non-trivial numerical computation. There has been a lot of work on bringing better storage and processing support for numerical computation in database systems (e.g., [1, 11, 16]). However, much of that work remains highly database-centric. First, users must learn another language, very often SQL or a variant that may be unfamiliar to them. Second, these systems force users to draw an explicit boundary between database processing and processing by the host programming language; where to draw this boundary is an optimization decision that arguably should not be left to inexperienced users. Therefore, the cost of embracing a database-centric approach remains prohibitive for ordinary users.

*This work is supported by the NSF DDDAS program under award CNS-0540347.

[1] We are unaware of any such libraries for R, but it should be possible to incorporate these libraries as R packages.

To have a practical impact on the majority of the users of R (or any other language for numerical computing), we believe that a better approach is to make it completely *transparent* to the users how we support efficient I/O. Transparency means no SQL, or any new language to learn. Transparency means that existing code should run without modification, and automatically gain I/O-efficiency.

Achieving transparency is challenging. First, besides making data layout and elementary operations I/O-efficient, what are the higher-level optimization opportunities and how critical are they? Second, what does it take to integrate I/O-efficient data layout, algorithms, and optimizations seamlessly into an existing language environment? R and MATLAB are interpreted; is it possible to implement the higher-level optimizations without switching to compilation instead? To what extent do we need to modify the existing implementation of these language environments?

This paper presents a vision for *RIOT* (*R with I/O Transparency*), a system that makes R I/O-efficient in a transparent fashion, without requiring users to learn a new language or rewrite their code. Although we are currently focusing on R, we expect many of our techniques to work for other numerical computing environments such as MATLAB. We will first report our experience of implementing *RIOT-DB*, a prototype system that uses a generic relational database system as a backend (while hiding it completely from users). Use of a relational database system allows us to explore both limitations and opportunities offered by the full range of database system features. Some features, such as the view facility, while seemingly unrelated to I/O-efficiency, turned out to be indispensable to RIOT-DB. We show that it is indeed possible to achieve I/O-efficiency transparently, not only because R already has high-level language constructs, but also because we are able to interface the host language environment and the database system in a clever way to enable high-level, inter-operator optimizations, such as avoiding materialization of intermediate results, deferring and reordering operations, etc. We demonstrate that, despite the overhead and inadequacy of generic database systems in handling array data and numerical computation, RIOT-DB's high-level optimizations enable it to significantly outperform R in many common scenarios. Finally, we outline our vision for the next generation of RIOT, based on our experience with RIOT-DB.

## 2 Related Work

Many relational database systems have introduced support for arrays (standardized in SQL99): e.g., Oracle's VARRAY and nested table, and PostgreSQL's ARRAY. Data cubes [3] can also be regarded as high-dimensional arrays, and many systems offer good storage and query support for them; however, these systems focus on OLAP-style queries instead of numerical computation over arrays. Furthermore, all these solutions fall under the database-centric approach discussed in Section 1, and therefore it is difficult to gain traction in scientific and statistical user communities.

There have been a number of database systems specialized in array processing; because of limited space, we describe only a few representatives here. RasDaMan [1] provides extensive support for multidimensional arrays with its own query language, RasQL, which extends SQL92. Queries are translated into an array algebra and optimized using a large collection of transformation rules. The storage manager utilizes various array tiling strategies to support different access patterns. AML [11] is another declarative language for manipulating arrays, along with a suite of query processing and optimization techniques. The system allows MATLAB users to issue AML queries and bring their results into MATLAB for further processing. Again, in contrast to our fully transparent approach, such systems can be considered database-centric (even though AML is not SQL-based), because users must explicitly draw a boundary between database processing and processing by the host programming language. Also, RasDaMan and AML target different application domains from ours and therefore do not treat operations such as matrix multiplication as first-class citizens; therefore, high-level optimizations involving these operations are difficult.

ASAP [16] is an array processing system that supports primitive operations oriented towards scientific computing. It also features ChunkyStore, a storage manager highly optimized for storing multi-dimensional arrays. The work in [16] focuses primarily on demonstrating the I/O-efficiency brought by ChunkyStore to individual operations; on the other hand, in this paper we emphasize more on the high-level, inter-operation optimizations. Also, we have the additional goal of making an existing language environment I/O-efficient in a transparent manner to users.

Lots of work from the scientific and high-performance computing communities has gone into developing I/O-efficient libraries (e.g., [18, 13]). Toledo gives an excellent survey [17] on out-of-core linear algebra algorithms. This line of work provides a solid foundation for us to build on. However, as we have mentioned in Section 1 and will show in the rest of this paper, higher-level optimizations are just as critical in ensuring I/O-efficiency; a library-only approach is insufficient by itself.

Finally, we note that there are many interesting connections between our techniques (often databases-inspired) and those from the programming languages community. Many of our optimizations have analogies in compiler research: e.g., deferred evaluation, forward substitution, loop fusion, and array contraction. However, there are also notable differences. On the highest level, work on programming languages tends to focus on improving performance of memory-resident programs and/or their parallelization; I/O issues are often not considered. Also, much of this line of work, further discussed below, assumes a compilation approach, while RIOT tries to work within the confines of R's interpretation approach.

Traditional compiler optimization techniques for array languages (e.g., Fortran 90) first translate array statements into scalar operations (expressed as loops), and then perform data dependence analysis and code transformations such as loop reversal and loop fusion. This approach, however, as Lewis et al. [9] pointed out elegantly, "solves the problem at a greater conceptual distance from the source of the problem and at a greater cost." Instead, RIOT optimizes at the higher level of array operations, an approach also used by [4, 6, 9, 8, 15].

Guibas and Wyatt [4] studied delayed evaluation in APL code compilation, i.e., deferring the computation of intermediate results in an APL expression until the moment they are needed. During evaluation, intermediate results are "streamed" in time, instead of being materialized in temporary arrays. Hwang et al. [6] generalized the idea and applied it to Fortran 90. They also support statement merges, whereby certain adjacent statements can be merged into one and processed as a single loop without materializing temporary arrays. Lewis et al. [9] perform dependence analysis among statements and identify clusters of statements that are "contractible" into a single loop. Joisha and Banerjee [8] studied how to minimize array storage in MATLAB, using program analysis to identify opportunities to reuse storage allocated for one array on another. Rosenkrantz et al. [15] performs inter-statement optimiza-tion to avoid materializing temporary arrays that can be obtained by "shuffling" other stored arrays. Although RIOT effectively performs similar optimizations as those cited above, our techniques are different due to the interpreted nature of R. Also, RIOT is far more aggressive in deferring evaluation (e.g., converting assignments to deferrable function evaluations) and performing optimization (e.g., considering different data layouts and algorithms, and reordering matrix multiplications) than these works.

Menon and Pingali [12] presented a framework for detecting high-level matrix operations written in loop-oriented codes. The codes are represented in an intermediate form, which is then optimized using heuristic rule-based transformations; RIOT considers a broader set of optimizations. Iu and Zwaenepoel [7] proposed a Java bytecode rewriting tool that automatically detects and converts compiled Java code that directly manipulates database tables as Java collections (instead of using JDBC) into more efficient SQL queries. Techniques from this line of work complement RIOT in the sense that we can apply them to programs written with lower-level loops (as opposed to high-level array operations) and then make them amenable to RIOT's optimizations.

With the recent interest in blurring the boundary between programming and query languages (e.g., declarative networking [10], Pig Latin [14], LINQ), we expect that more and more connections between programming languages and databases communities will become relevant to this research.

# 3 Opportunities for Improving R

To illustrate sources of I/O-inefficiency in R and opportunities for improvement, consider the following example.

**Example 1.** *We are given a large number of points in a* 2*-d space, whose coordinates are stored by vectors* `x[1:n]` *and* `y[1:n]`*. Given two other points* `(xs,ys)` *and* `(xe,ye)`*, we want to compute the lengths of paths between them via each of the points given earlier. We then draw* 100 *such lengths at random. The following R code accomplishes this task. Note that most operations in R are vectorized (e.g.,* `^2` *squares every element of a vector and returns the results in a new vector).*

```
(1)  d <- sqrt((x-xs)^2+(y-ys)^2) + sqrt((x-xe)^2+(y-ye)^2)
(2)  s <- sample(length(x),100) # draw 100 samples from 1:n
(3)  z <- d[s] # extract elements of d whose indices are in s
```

**Avoiding Intermediate Results** It is common for an expression to involve multiple operations, such as Line (1) of Example 1. R would generate an intermediate result for each of these operations: first `x-xs`, then `(x-xs)^2`, and so on. If memory can hold all data objects including intermediate results, there is no problem. However, when intermediate results accumulate and leave insufficient memory, thrashing can occur. Consider Line (1) again. R would generate a total of twelve intermediate results, all vectors of length `n`. Even with a smart garbage collector that immediately reclaims memory as soon as an intermediate result is no longer needed, there can be multiple intermediate results alive at the same time. When evaluating `(y-ye)^2`, for example, three intermediate results are alive: `sqrt((x-xs)^2+(y-ys)^2)`, `(x-xe)^2`, and `(y-ye)`. Together with `x` and `y`, we have five `n`-vectors that can easily cause thrashing if `n` is large.

If we were to hand-code Line (1), we could in fact compute `d` without materializing any of the twelve intermediate results, by using an explicit loop over `1:n` and computing one element of `d` at a time. This strategy would require a negligible amount of memory beyond the two inputs and the output. The question, of course, is how we can accomplish this optimization automatically.

**Deferred and Selective Evaluation** A closer examination of the R code in Example 1 reveals that not all elements of `d` need to be computed; in fact, only 100 of them are eventually used on Line (3). Nonetheless, R will happily compute the entire `d`, wasting both computation and I/Os. If we could somehow defer the evaluation of `d` until we know which 100 elements are needed, we would selectively compute them by accessing the corresponding elements in `x` and `y`. With this optimization, we would reduce the cost of accessing `x` and `y` (especially if they had been swapped out previously), and even avoid materializing the named object `d`.

One might argue that a programmer should know enough to avoid useless computation, but such code is not uncommon for those without formal training in programming. Our hope is that a programmer can focus on *what* they want to compute instead of *how* they want to compute it, and leave the rest to RIOT.

**Example 2.** *Given three matrices* `A` *(with dimensions $n_1 \times n_2$),* `B` *($n_2 \times n_3$), and* `C` *($n_3 \times n_4$), we want to compute* `A %*% B %*% C`*, their product expressed in the syntax of R. R would first multiply* `A` *and* `B`*, and then multiply the result and* `C`*.*

*Internally, R implements matrix multiplication as follows, where* `T` *denotes the result of* `A %*% B`*. This algorithm performs a total of $n_1 n_2 n_3$ multiplications. R by default uses a* column layout *for matrices; i.e., elements are stored in the column-major order.*

```
for (j in 1:n3)
  for (i in 1:n1) {
    T[i,j] <- 0
    for (k in 1:n2)
      T[i,j] <- T[i,j] + A[i,k]*B[k,j]
  }
```

**Optimizing Data Layout and Algorithms** When the size of data exceeds the memory capacity, data has to be swapped in and out of memory in blocks. For efficiency, how we lay out data should correspond to how we access it, so that each disk block we read will bring in a maximum amount of useful data. The data access pattern of matrix multiplication in Example 2 is significantly more complex than the sequential access pattern for most vector operations in Example 1. Therefore, a closer look is warranted.

Suppose the size of a disk block is $B$, and the size of available memory is $M$, where $M \ll \min\{n_1 n_2, n_2 n_3, n_3 n_4\}$. In the algorithm of Example 2, to compute each column of `T`, we must access one column of `B`, and the entire `A` in row-major order. If both `A` and `B` use column layout (R default), each access to `A` would likely result in a page fault, bringing the total I/O cost to a huge $\Theta(n_1 n_2 n_3)$. Had we been smarter to choose row layout for `A`, the total I/O cost would have been reduced to $\Theta(n_1 n_2 n_3 / B)$.

Higher I/O-efficiency can be gained by further tweaking the access pattern of matrix multiplication. Borrowing the idea from block nested-loop join, we could read as many rows of `A` as possible into memory while leaving enough memory to update the corresponding rows of `T` and a block to scan `B` in column-major order. The total I/O cost would be reduced to $\Theta(\frac{n_1 n_2 n_3 (n_2 + n_3)}{BM})$. As we will show later, however, there are even better strategies if we move beyond the built-in row and column layouts supported by R.

**Reordering Computation** R computes a chain of matrix multiplications in the order specified by the program. For Example 2, this strategy requires $n_1 n_2 n_3 + n_1 n_3 n_4$ multiplications and a commensurate number of I/Os. However, noting that matrix multiplications are associative, we could instead compute `A %*% (B %*% C)`, which would require $n_2 n_3 n_4 + n_1 n_2 n_4$ multiplications. Depending on the values of $n_1, \ldots, n_4$, reordering the multiplication may significantly reduce both computation and I/O. The challenge is to let RIOT make such optimization decisions, much in the same way as a database query optimizer.

# 4 RIOT-DB: Database as a Solution?

Having discussed some sources of I/O-inefficiency in R, we now describe our experience of implementing RIOT-DB, a prototype system that addresses these sources of inefficiency using a relational database system as a backend. While doing so, RIOT-DB maintains complete transparency; i.e., existing R programs can benefit from RIOT-DB without any modification. We are aware of the overhead and inadequacy of relational database systems for this task, as shown by previous work; ASAP, for example, revealed gross inefficiency of such systems at the storage level [16]. We still chose this option because it enabled rapid prototyping and offered an opportunity to investigate not only the limitations but also the potential of leveraging other relational database features (e.g., views, query optimization) in a new context.

**Interfacing with R** Instead of rebuilding R from scratch to make it I/O-efficient, we decided on a minimally invasive approach. We would build RIOT-DB as an R package using R's extensibility features, and avoid modifying the core R code whenever possible. RIOT-DB can be dynamically plugged into an R environment, and immediately adds I/O-efficiency to R programs. The decision to be modular and minimally invasive is important, since we plan to apply our techniques to other environments (e.g., MATLAB).

RIOT-DB defines three new data types, `dbvector`, `dbmatrix`, and `dbarray` (with an arbitrary number of dimensions), which correspond to R's built-in `vector`, `matrix`, and `array`. The new types implement the same interfaces as their built-in counterparts. Users do not need to know whether an object they are dealing with has a RIOT-DB type or a built-in type. R's *generics* mechanism [2] enables this transparency. Briefly put, a generic function in R is associated with a collection of concrete methods which share the same formal arguments, but differ in the classes of the arguments. When a call to a generic function is evaluated, a method is selected according to the classes of the actual arguments. This is analogous to method overloading in some object-oriented programming languages like C++.

We illustrate how to define a new class and to use the generics mechanism by an example. Suppose we want to add an + operator for the new `dbvector` class. Below are the steps we follow.

1. Define a new class for `dbvector` by: `setClass("dbvector",representation(size="numeric",···))`, where `representation(···)` defines the names and types of `dbvector`'s members.

2. Define a method for adding two `dbvectors` and register it with the + generic method:

```
setMethod("+", signature(e1="dbvector", e2="dbvector"),
  function(e1,e2) {
    .Call("add_dbvectors", e1, e2)
  }
)
```

   The above code specifies that when the two operands of the + call are both of `dbvector` type, the provided function should be invoked. The provided function further calls a C function `add_dbvectors`. Notet that `.Call` can call C functions at runtime from dynamically loaded libraries (.dll on Windows platform or .so on Unix-like platforms).

3. Implement the addition logic in a C function:

```
SEXP add_dbvectors(SEXP e1, SEXP e2) {
  /* implementation */
}
```

4. When two `dbvector` objects are added in user R code, e.g., `a+b`, all functions registered with the + generic will be checked for argument type match. The result is that our function in Step 2 is selected and eventually our custom C code is executed.

Fortunately, the object-oriented programming facility as illustrated above is not peculiar to R. Many other numerical computing environments, such as MATLAB, also provide mechanisms for registering new classes and overloading operators. Thus, we expect our general design to be portable to other environments.

**A Strawman Solution** A straightforward way for RIOT-DB to leverage a relational database system is to map every RIOT-DB object to a database table. A `dbvector` object would be mapped to a table with schema $(\mathtt{I}, \mathtt{V})$, where the primary key `I` stores an index, and `V` stores the corresponding vector element. An $n$-dimensional `dbarray` object would be mapped to schema $(\mathtt{I}_1, \ldots, \mathtt{I}_n, \mathtt{V})$, where the array indexes $(\mathtt{I}_1, \ldots, \mathtt{I}_n)$ serve as the primary key. The function `add_dbvectors`, which adds two `dbvector` objects corresponding to tables `E1` and `E2`, would compute the result using the following SQL query:

```
SELECT E1.I, E1.V+E2.V AS V FROM E1, E2 WHERE E1.I=E2.I
```

The result of the above query would be stored in another database table, which is then associated with the `dbvector` object representing the result of addition. Execution of this query would carry a very small memory footprint.

As shown in [16], however, storing array indexes in tables incurs significant storage and processing overhead, which grows linearly with the number of dimensions. Database query processing also carries overhead, and usually cannot match R's raw performance.

Some of these problems will go away if we move to a more specialized database system that uses, for example, a smarter storage manager like ChunkyStore [16]. However, deeper issues still remain. In particular, this strawman approach leverages the power of a database system only at an intra-operation level, and fails to address any I/O-inefficiency that exists at the inter-operational level. One could argue that R is partly to blame because RIOT-DB can only take control of individual operations involving RIOT-DB types. Next, we show how to enable higher-level optimizations within the confines of this interface between R and RIOT-DB.

## 4.1 Towards Inter-Operation Optimization

Interestingly, views provide a natural mechanism for RIOT-DB to tap more into database systems' advanced features through its limited interface with R. Recall that when we define a view using a query (over tables or other views), the system simply records this query without evaluating it. When the system executes a query involving a view, the query is expanded by replacing references to the view by its definition query.

We map each RIOT-DB object to a database table or view. The result of operating on RIOT-DB objects becomes a view, whose definition encapsulates the computation involved in generating this result. However, no computation actually takes place (yet). For example, RIOT-DB's `add_dbvectors` function simply defines the following view to capture the result of adding two `dbvector` objects, and associates the view with the result object:

```
CREATE VIEW E3(I,V) AS
SELECT E1.I, E1.V+E2.V FROM E1, E2 WHERE E1.I=E2.I
```

For a complex R expression such as Line (1) of Example 1, RIOT-DB would define one view for each intermediate result object. The view definition for d, when expanded by the database system, would look like the following:

```
CREATE VIEW D(I,V) AS
SELECT TMP1.I, TMP1.V+TMP2.V
FROM (SELECT I, SQRT(V) AS V
      FROM (SELECT TMP3.I AS I, TMP3.V+TMP4.V AS V...))  TMP1,
     (SELECT I, SQRT(V) AS V
      FROM (SELECT TMP5.I AS I, TMP5.V+TMP6.V AS V...))  TMP2
WHERE TMP1.I=TMP2.I
```

In effect, the view mechanism allows RIOT-DB to build up, one operation at a time, bigger and bigger view definitions that correspond to more and more complex R expressions. With a view representing a complex, multi-operation expression, we are now ready to unleash other features of database systems.

**Avoiding Intermediate Results** To compute the result of a complex R expression, RIOT-DB evaluates the definition query of the corresponding view. Most database systems optimize and compile the query into a tree-shaped plan, and use an *iterator-based* model to execute it. Query execution proceeds in a recursive fashion, where each plan operator obtains its input tuples one at a time, as needed, from its child operators. Leveraging this execution model, RIOT-DB effectively pipelines processing among plan operators, and eliminates the need to materialize intermediate results (although in some cases the database system can still decide to materialize for performance). Compared with the strawman approach, RIOT-DB avoids storing intermediate results in temporary tables on disk. Compared with plain R, RIOT-DB avoids creating large intermediate results in memory; in a memory-constrained setting, I/O savings resulted from fewer virtual memory swaps can be substantial.

For example, to compute d in Example 1, RIOT-DB only needs a single pass over the tables associated with x and y, and incurs no additional I/Os for intermediate results.

**Deferred and Selective Evaluation** RIOT-DB does not restrict the use of views to unnamed intermediate results produced within a single R expression. Named objects can be created with views as well,[2] effectively deferring their evaluation. If a named object is subsequently referenced, RIOT-DB simply uses the view associated with that object. In Example 1, object z on Line (3) would correspond to the following view (note how dereferencing a vector with a vector of indices translates cleanly to a join between them):

```
CREATE VIEW Z(I,V) AS
SELECT S.I, D.V FROM D, S WHERE D.I=S.V
```

At this point, RIOT-DB has defined view D, but not yet computed its content. Suppose we now want the result in z. RIOT-DB will effectively compute the following query, where *xs*, *xe*, *ys*, and *ye* should be replaced with their actual values:

```
SELECT S.I, SQRT(POW(X.V-xs,2)+POW(Y.V-ys,2))
          + SQRT(POW(X.V-xe,2)+POW(Y.V-ye,2))
FROM X, Y, S WHERE X.I=Y.I AND X.I=S.V
```

Since S is very small, a reasonable database query optimizer would pick an index nested loop plan, which probes X and Y with each S.V value and computes the SELECT clause. Hence, RIOT-DB is able to compute just those d elements that are actually used, thereby saving both computation and I/O.

[2]There is one technicality here. Assignments introduce dependencies on views created by RIOT-DB, but R does not notify RIOT-DB of assignment operations. To be able to safely drop views, RIOT-DB must track such dependencies. Therefore, we had to introduce this additional hook for R assignments (which was the only modification we made to the core R code).

**Optimizing Algorithms and Data Layout** Although generic database systems handle vectors reasonably well, optimizations beyond vectors are still worrisome. For example, RIOT-DB defines matrix multiplication (cf. Example 2) as follows:

```
SELECT A.I, B.J, SUM(A.V*B.V) AS V
FROM A, B WHERE A.J=B.I GROUP BY A.I, B.J
```

The optimized query plan[3] does a hash join on A.J=B.I, and then sorts the result by (A.I, B.J) to perform group-by and aggregation. Unfortunately, as we will show in Section 5, this plan is far from the optimum. We believe the problem lies in that SQL is too low-level for representing many linear algebra operations; optimizing at this level is much less effective than if we know the high-level semantics of these operations. We will revisit this point when presenting our design for the next generation of RIOT in Section 5.

In terms of matrix data layout, there is little in a generic database system for RIOT-DB to leverage. RIOT-DB can specify either row or column layout by changing the order of index attributes in the primary key. With automated database design techniques, it might be possible to automate the choice between these two layouts. However, it is awkward to specify more advanced layouts such as *tiling* (further discussed in Section 5) in a generic database system, let along finding the best tiling strategy.

## 4.2 Experiment Results

To evaluate the performance of RIOT-DB and the savings obtained by its various optimizations, we compare four approaches:

- **Plain R**;
- **RIOT-DB/Strawman**, as described earlier in this section (with no views);
- **RIOT-DB/MatNamed**, which uses views, but materializes all named objects;
- **RIOT-DB** (the full version).

Our RIOT-DB uses MySQL with MyISAM storage engine as the backend. All experiments are conducted on a Solaris 10 machine with an AMD Opteron 275 processor. To limit the burden placed on the testing machine, we did not test with very large vectors whose sizes are greater than the actual amount of physical memory. Instead, we simulated a limited-memory environment by using a simple program to lock down a large portion of the memory. The program uses the shmat(2) system call on Solaris, with the SHM_SHARE_MMU flag. This operation has the consequence of locking down the allotted memory pages in physical memory such that they will never be paged out.

We run the code in Example 1, capping the available amount of physical memory at 84MB, just enough to hold the R runtime plus two vectors with $2^{22}$ elements each. To be fair, we set the same memory cap for RIOT-DB variants (which include the MySQL sever). To force computation of z, we add a final line: print(z).

Two metrics are used to compare the performance of the four approaches: execution time and disk I/O. To measure the amount of I/O, we utilize the DTrace facility on the Solaris platform. From *Solaris Dynamic Tracing Guide*, "DTrace is a comprehensive dynamic tracing facility that is built into Solaris that can be used by administrators and developers on live production systems to examine the behavior of both user programs and of the operating system itself." We use DTrace to monitor different statistics for plain R and RIOT-DB:

[3]Although we implemented RIOT-DB with MySQL, we obtained this plan on a commercial database system with a well-regarded optimizer.

- For plain R, I/O is caused by the swapping of data into and out of the physical memory. We thus monitor virtual memory paging statistics.

- For RIOT-DB, virtual memory paging activity is negligible assuming there is enough memory to run R and MySQL; most I/O is caused by the MySQL database server reading and writing its data and index files. Therefore we monitor disk I/O statistics pertaining to MySQL files.

Results are shown in Figure 1. When vectors are sized at $2^{21}$ and $2^{22}$, RIOT-DB/Strawman underperforms plain R, even though R already suffers significantly from thrashing. The reason lies in the overhead of MySQL for storage (additional columns for array indexes) and numerical computation, and in that MySQL is used only at an intra-operation level. Intermediate results are particularly damaging to RIOT-DB/Strawman, because it writes them all into database tables. Nonetheless, MySQL-managed I/Os are mostly bulky and sequential, and therefore do not impact the execution time as much as the virtual memory I/Os incurred by R; also, performance of RIOT-DB/Strawman degrades linearly with the data size, much more gracefully than plain R.

Once we enable inter-operation optimizations, performance improves dramatically. RIOT-DB/MatNamed, by avoiding materialization of nameless objects and pipelining query execution, already nets significant gains over R. RIOT-DB is barely visible in the figures because it is so much faster than others. By further deferring evaluation across statements and using the database optimizer to avoid unnecessary evaluation, it is able to outperform plain R by orders of magnitude. These results demonstrate the importance and potential of inter-operation optimizations.

## 5 RIOT: The Next Generation

RIOT-DB has shown the feasibility of bringing I/O-efficiency to R in a transparent manner, revealed the overhead and inadequacy of generic database systems for numerical computation, and demonstrated the potential of higher-order, database-style optimizations. Based on these lessons, we now outline our design for the next generation of RIOT, which is currently under development.

**Data Storage and Layout Options** To address well-understood inefficiencies of the simple relational representation for arrays, RIOT draws ideas from ChunkyStore [16]: e.g., no explicit storage of array indices, and flexible *tiling* (called *chunking* in ChunkyStore). With tiling, an array is partitioned into (hyper)rectangular *tiles*; each tile is stored in a disk block, but the aspect ratio of tiles can be controlled. For matrices, row and column layouts correspond to tiling strategies where tiles are long and skinny; however, higher I/O-efficiency is possible with more flexible tiling strategies.

Beyond the features of ChunkyStore, RIOT also provides advanced *linearization* options for controlling the order in which tiles are stored on disk. This ordering is important because of the performance difference between sequential and random I/O. RIOT plans to support linearizations based on space-filling curves, for arrays whose access patterns are not known in advance.

**Expression Algebra** RIOT-DB made extensive use of SQL views to piece together individual operations into larger expression for optimization and execution. However, we do not need SQL views per se for this purpose. With an appropriate high-level expression algebra, RIOT can build up an expression DAG, operation by operation, in the same fashion as RIOT-DB.

RIOT's expression algebra includes standard linear algebra operations, such as matrix multiplication and LU decomposition. This approach departs from those that are more *minimalist* in design (e.g., RasDaMan [1], AML [11]), where such operations are expressed using a combination of lower-level operators. From the RIOT-DB experience, we have seen that a minimalist approach might make high-level optimization more difficult (e.g., if matrix multiplication is expressed in relational query operators).

Given the abundance of previous work on array algebras and limited space, we will not go into the details of the RIOT algebra. Instead, we present one example to highlight some of the unique aspects of our design. Consider the following R code fragment. It computes vector `b` as the element-wise square of `a`, sets all elements of greater than 100 to 100, and then prints the first 10 elements:

```
b <- a^2; b[b>100] <- 100; print b[1:10]
```

Recall from Section 4 that we use views to defer evaluation as much as possible. However, as soon as an object is modified (e.g., `b[b>100] <- 100`), RIOT-DB forces computation and materialization of the object (in this case, `b`) before modifying it.

In RIOT, however, we propose to defer modifications as well, by modeling them using an operator that takes the old object state and the new content as input, and returns the new state as output. The expression `b[1:10]`, at the end of the code fragment, would be represented by the expression DAG shown in Figure 2(a). The `[]<-` operator models vector modification. Turning modifications (with side-effects) into functions (without side-effects) allows RIOT to defer evaluation further, build up larger expressions to optimize, avoid materializing modified objects, and, as we will see shortly, potentially eliminate the need to perform some modifications.

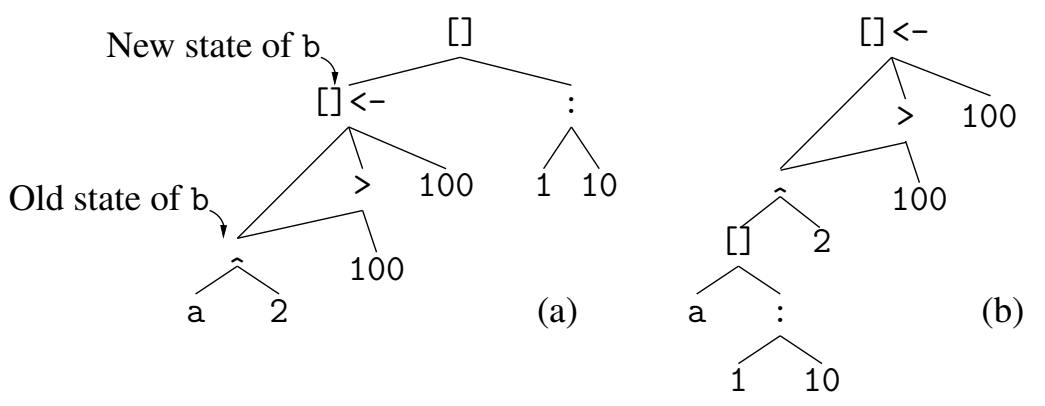

**Figure 2: Expression DAGs for `b[1:10]`.**

**Optimization** Given an expression DAG consisting of high-level operators, RIOT carries out database-style optimizations. For example, via a series of transformation rules, RIOT can optimize the expression DAG in Figure 2(a) into the one in Figure 2(b). Note that the "selection" of the first 10 elements in modified `b` has been "pushed down" all the way onto `a`. Hence, modifications to `b` (as well as tests of whether an element of `b` should be modified) only need to be executed on 10 elements.

For examples of RIOT optimizations specific to numerical computation, we show how to optimize the matrix multiplication chain in Example 2. First, consider multiplying two matrices `A` $(n_1 \times n_2)$ and `B` $(n_2 \times n_3)$ with memory $M$ and block size $B$. Previous work has established a lower bound [5, 17] of $\Omega(n_1 n_2 n_3 / (B\sqrt{M}))$ I/Os for algorithms requiring $\Theta(n_1 n_2 n_3)$ multiplications,[4] and proposed an algorithm with matching complexity [17], which we describe below. We present an alternative proof for this lower bound in Appendix A that gives more insight into how to achieve it, which leads naturally to the algorithm.

This algorithm divides available memory into three equal parts, each storing a $p \times p$ (where $p = \sqrt{M/3}$) square submatrix: one from A, one from B, and the other for the result T. For simplicity, assume that $n_1$, $n_2$, and $n_3$ are multiples of $p$. The algorithm, captured by the pseudocode below, looks similar to the one in Example 2, except that it operates on the level of submatrices:

```
for (i in 1:(n1/p))
  for (j in 1:(n3/p)) {
    Tsub <- matrix(0,p,p)
    for (k in 1:(n2/p)) {
      read Asub <- A[(i*p-p+1):(i*p),(k*p-p+1):(k*p)] from disk
      read Bsub <- B[(k*p-p+1):(k*p),(j*p-p+1):(j*p)] from disk
      Tsub <- Tsub + Asub %*% Bsub
    }
    write Tsub as T[(i*p-p+1):(i*p),(j*p-p+1):(j*p)] to disk
  }
```

[4]There exist algorithms that use fewer multiplications, such as Strassen's, although they are harder to implement and numerically less stable.

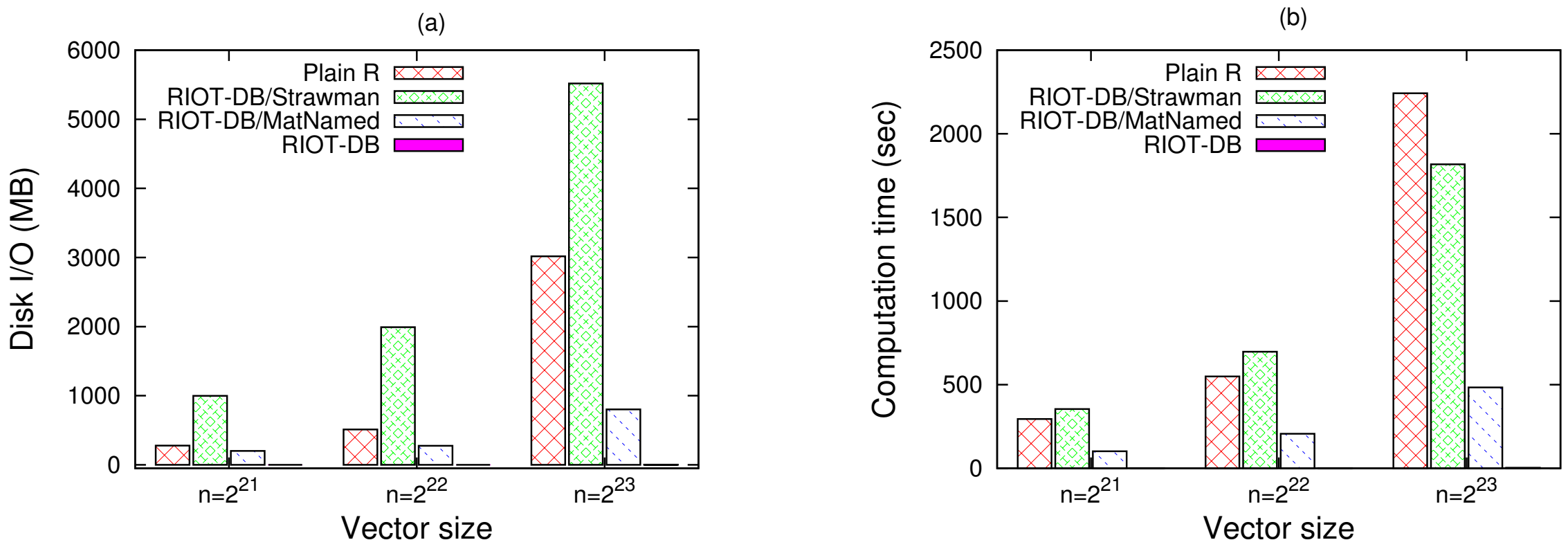

**Figure 1: Performance of R vs. RIOT-DB variants.**

A storage layout strategy that works well with this algorithm is to use square tiles of area $B$, such that each $p \times p$ submatrix requires $O(p^2/B)$ I/Os. The total number of I/Os is $\Theta(n_1 n_2 n_3/(B\sqrt{M}))$, matching the lower bound. For large matrices, this algorithm beats the one in Section 4 inspired by block nested-loop join, which uses row and column layouts.

Stepping up a level, we now discuss how to optimize a chain of matrix multiplications. With dynamic programming [5], we can find a multiplication order that minimizes the total number of multiplications. Let $N$ denote this number. Using the matrix multiplication algorithm above and square tiling for all matrices, RIOT can compute the chain in $\Theta(N/(B\sqrt{M}))$ I/Os. Asymptotical optimality is shown in Appendix B.

To illustrate the effectiveness of RIOT optimizations related to matrix multiplications, we compare four strategies for computing `A %*% B %*% C`:

- **RIOT-DB** uses a plan consisting of two hash-join-sort-aggregate subplans (Section 4), one to first multiply A and B, and the other to multiply with C;
- **BNLJ-Inspired** assumes that the matrices use row, column, and column layouts respectively, and performs the matrix multiplications in order, using the algorithm in Section 4 inspired by block nested-loop join;
- **Square/In-Order** assumes square tiling for all matrices, and performs the matrix multiplications in order, using the algorithm described in this section;
- **Square/Opt-Order** also employs square tiling and the same matrix multiplication algorithm as Square/In-Order, but first uses dynamic programming to find the best multiplication order.

Suppose A, B, and C have dimensions $n \times \frac{n}{s}$, $\frac{n}{s} \times n$, and $n \times n$, respectively, where $s > 1$ is a skewness factor, which causes Square/Opt-Order to choose the multiplication order A(BC). The block size $B = 1024$. Figure 3(a) compares the calculated I/O costs of the four strategies[5] for $n = 100000$ and 120000, and for memory sizes of 2GB and 4GB. We see a progression of improvements as more optimizations are introduced, and this trend is consistent for all parameter settings tested. Figure 3(b) shows the results when we vary $s$, the skewness factor. Memory is set at 2GB and $n = 100000$. RIOT-DB is no longer shown because it performs far worse than others. As $s$ increases, the performance gap between Square/Opt-Order and others widens, demonstrating the importance of optimizing the multiplication order.

**Discussion** This section has sampled some important ideas that we plan to investigate in RIOT. There are some other interesting research issues, such as how to make better materialization and data layout decisions. Briefly, RIOT needs materialization to complement deferred evaluation; otherwise, RIOT may have to repeat the same computation across multiple complex expression DAGs that it has built up. RIOT needs to jointly optimize data layout and processing, and consider the option of dynamically changing data layout should access patterns change. These decisions are challenging because RIOT must make them on the fly, without knowing what computation might come in the future. We believe there is an opportunity for developing better, practical solutions by combining ideas from scientific computing, programming languages, and databases.

## 6 Conclusion

I/O-efficiency and query optimization are features standard to modern database systems, but sorely missed by many numerical computing environments like R. Instead of forcing R users to learn to use database systems, however, we propose to build a system called RIOT that brings I/O-efficiency and database-style optimizations in a completely transparent manner to R users.

Our experience of implementing RIOT-DB has provided many insights, the most important of which is the fact that I/O-efficiency cannot be achieved without high-level, inter-operation optimizations. Even though the generic database system used by RIOT-DB carries enormous overhead in storing and processing arrays, its pipelined execution model and query optimizer are able to turn the tide to its favor in many cases. With a specialized storage engine, algorithms, and database-style optimization strategies tailored towards numerical computing, we expect the next generation of RIOT to make significant further gain in I/O-efficiency. Another pleasant surprise from this experience is that transparency is indeed possible, with ideas such as deferred evaluation and clever mechanisms to implement them within the confines of R. As future work, we plan to investigate how to apply our techniques to other language environments beyond those intended for numerical computing.

[5] To focus our comparison on evaluation strategies, Figure 3 excludes the overhead of storing array indexes in RIOT-DB; this adjustment has no effect on the relative ordering of performance, however.

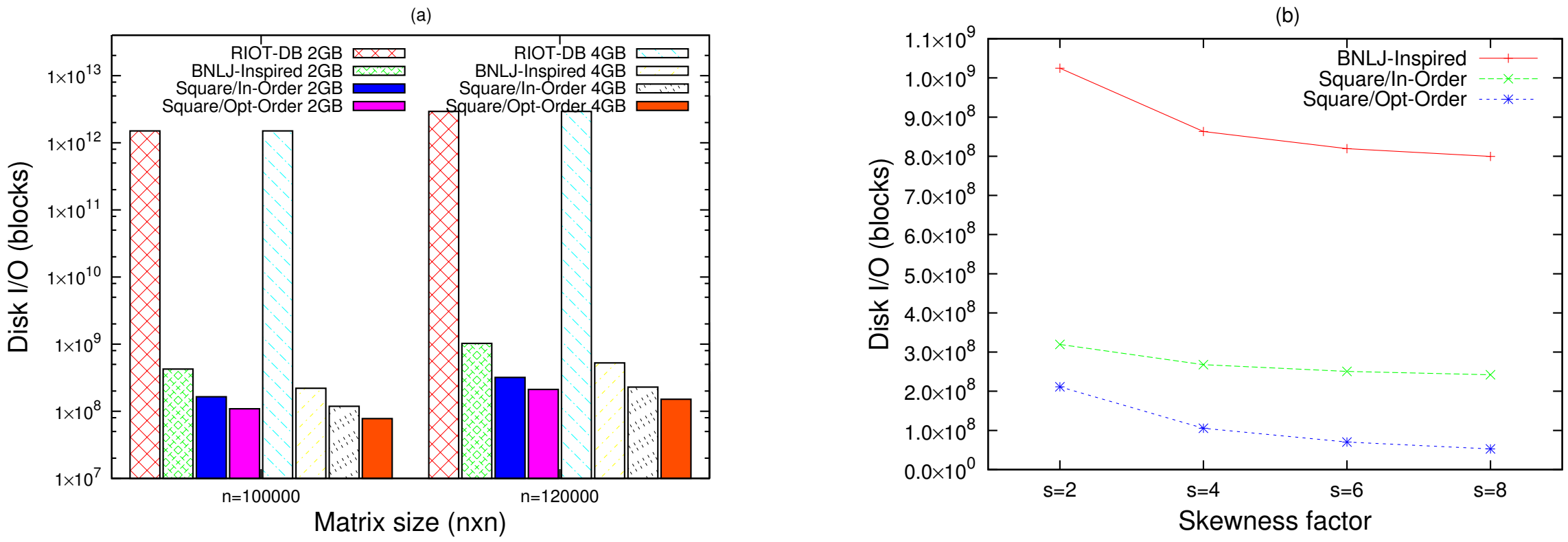

**Figure 3: I/O costs of a chain of three matrix multiplications.**

# APPENDIX

## A I/O Lower Bound of Matrix Multiplication

**Problem:** Given two matrices $\mathbf{A}(m \times l)$ and $\mathbf{B}(l \times n)$, compute their product $\mathbf{C} = \mathbf{AB}$. The available memory can hold $M$ scalar numbers. Suppose $\min\{ml, ln, mn\} \gg M$. $\mathbf{A}$ and $\mathbf{B}$ initially reside on disk. Each disk block can store $B$ numbers. Assuming any algorithm requiring $\Theta(lmn)$ scalar multiplications can be used, give an optimal schedule that minimizes the amount of I/O in terms of disk blocks read/written.

**Solution:** We first give a lower bound for the amount of I/O and then give a schedule that achieves this lower bound.

At any time, the memory contains elements from $\mathbf{A}$, $\mathbf{B}$ or $\mathbf{C}$. During $M$ number of I/Os, the number of distinct elements from the three matrices must be $\leq 2M$ (elements initially in memory plus newly fetched ones). Let the number of distinct, *active* elements from $\mathbf{A}$, $\mathbf{B}$, and $\mathbf{C}$ that contribute to the matrix multiplication be $a, b$ and $c$, respectively. An active element is one that either participates in element multiplication (elements in $\mathbf{A}$ and $\mathbf{B}$), or gets assigned (elements in $\mathbf{C}$). Further suppose that the $a$ elements from $\mathbf{A}$ are taken from $U$ rows, each having $u_i$ elements, where $i = 1, \ldots, U$. Similarly, suppose that the $b$ elements from $\mathbf{B}$ are taken from $V$ columns, each having $v_j$ elements, where $j = 1, \ldots, V$. Note that if active elements from row $i$ of $\mathbf{A}$ and column $j$ of $\mathbf{B}$ are multiplied and contribute to $\mathbf{C}[i, j]$, then $\mathbf{C}[i, j]$ must appear in memory. Thus we have

$$
\begin{aligned}
a &= \sum_{1 \leq i \leq U} u_i \\
b &= \sum_{1 \leq j \leq V} v_j \\
c &\geq UV \\
a + b + c &\leq 2M.
\end{aligned}
$$

Now consider the number of scalar multiplication operations, which we denote by $Z$, that can be performed with the above constraints. Note that $\mathbf{A}[h, i]$ and $\mathbf{B}[j, k]$ are multiplied if and only if $i = j$. Thus, for each row $i$ of $\mathbf{A}$ and each column $j$ of $\mathbf{B}$, the maximum number of scalar multiplications is $\min(u_i, v_j)$. Therefore,

$$
Z = \sum_{1 \leq i \leq U} \sum_{1 \leq j \leq V} \min(u_i, v_j).
$$

We want to maximize $Z$ in order to minimize the amount of I/Os. Without loss of generality, let us assume $u_1 \leq u_2 \leq \cdots \leq u_U$ and $v_1 \leq v_2 \leq \cdots \leq v_V$. Let all $u_i$'s that are greater than $v_1$ and less than $(v_1 + v_V)/2$ be $D = \{u_d, \ldots, u_{d+r}\}$, and all $u_i$'s that are between $(v_1 + v_V)/2$ and $v_V$ be $E = \{u_e, \ldots, u_{e+s}\}$. Let all $u_i$'s greater than $v_V$ be $F = \{u_f, \ldots, u_{f+t}\}$. Now consider the following change: We remove $v_1$ and $v_V$ and introduce two identical numbers $(v_1 + v_V)/2$. There are still $V$ numbers, with their sum unchanged. Consider the resulting $Z$. All elements in $D$ now contribute to $Z$ once because of the change from $v_1$ to $(v_1 + v_V)/2$. All elements in $E$ have their contribution to $Z$ decremented by one because of the change from $v_V$ to $(v_1 + v_V)/2$. So the net change in $Z$ is

$$
\begin{aligned}
\Delta Z &= \sum_{u \in D} u - \sum_{u \in E} u - v_1(|D| + |E| + |F|) - v_V|F| \\
&\quad + 2\frac{v_1 + v_V}{2}(|E| + |F|) \\
&= \left( \sum_{u \in D} u - v_1|D| \right) + \left( v_V|E| - \sum_{u \in E} u \right) \\
&= \sum_{d \leq i \leq d+r} (u_i - v_1) + \sum_{e \leq i \leq e+s} (v_V - u_i) \\
&\geq 0.
\end{aligned}
$$

If we repeat this operation, i.e., replacing the smallest and the largest numbers with two copies of their mean, we can eventually make all $v_i$'s equal. The same procedure can be done to $u_i$'s. Thus, when $Z$ achieves its maximum, we have $u_1 = \cdots = u_U = a/U$ and $v_1 = \cdots = v_V = b/V$. Now consider maximizing $Z$ under this condition. Let $u = a/U$ and $v = b/V$. Without loss of generality, we assume $u \leq v$, or equivalently $aV \leq bU$, in which case $Z$ becomes

$$
Z = UV \cdot u. \tag{1}
$$

We want to maximize (1) subject to:

$$
\begin{aligned}
u &\leq v \\
uU + vV + UV &\leq 2M
\end{aligned}
$$

Let $f = UVu + \alpha(v - u) + \beta(2M - uU - vV - UV)$. Using Lagrange multiplier, we know $Z$ gets its maximum when $\nabla f = 0$; that is:

$$
\frac{\partial f}{\partial u} = UV - \alpha - \beta U = 0 \tag{2}
$$

$$
\frac{\partial f}{\partial v} = \alpha - \beta V = 0 \tag{3}
$$

$$
\frac{\partial f}{\partial U} = uV - \beta u - \beta V = 0 \tag{4}
$$

$$
\frac{\partial f}{\partial V} = uU - \beta v - \beta U = 0 \tag{5}
$$

$$
\frac{\partial f}{\partial \alpha} = v - u = 0 \tag{6}
$$

$$
\frac{\partial f}{\partial \beta} = 2M - uU - vV - UV = 0 \tag{7}
$$

By (4), (5), and (6) we have $(u - \beta)V = (u - \beta)U$. Note that we cannot have $u = \beta$, because plugging it into (4) would yield $u = \beta = 0$. So we must have $U = V$. Now combining with (2) and (3), we get $U = 2\beta$. Plugging this into (4), we get $u = U$. Coming back to (7), we have $a = uU = \frac{2}{3}M$. We conclude that $Z \leq (\sqrt{2M/3})^3$.

Recall that $Z$ is the number of scalar multiplications performed during $M$ elements of I/Os. Because the algorithm for computing $C$ requires a total of $lmn$ scalar multiplications, the total number of element I/Os should be at least

$$
\frac{lmn}{\left(\sqrt{\frac{2M}{3}}\right)^3} M = \Theta\left(\frac{lmn}{\sqrt{M}}\right).
$$

In the best case, we can service these element I/Os with $\Theta\left(\frac{lmn}{B\sqrt{M}}\right)$ number of block I/Os, which is the lower bound.

The design of a schedule that achieves the minimum now becomes straightforward. The condition that yields the minimum is $a = b = c$ and $U = u_1 = \cdots = u_U = V = v_1 = \cdots = v_V$. We can divide $\mathbf{A}$, $\mathbf{B}$ and $\mathbf{C}$ into submatrices of size $p \times p$, where $p = \sqrt{M/3}$. The memory can hold exactly three such submatrices. For each block $\mathbf{C}_{i,j}$ in $C$, we perform the block matrix algorithm by loading and multiplying pairs of submatrices from $\mathbf{A}$ and $\mathbf{B}$ in turn—$\{\mathbf{A}_{i,1}, \mathbf{B}_{1,j}\}$,...,$\{\mathbf{A}_{i,k}, \mathbf{B}_{k,j}\}$—to compute $\mathbf{C}_{i,j}$. For each $\mathbf{C}_{i,j}$, we read in $\frac{2p^2}{B}\frac{l}{p}$ blocks, and write out the result, which has $\frac{p^2}{B}$ blocks. There are $\frac{mn}{p^2}$ submatrices in $\mathbf{C}$. So the total number of I/Os (blocks) is

$$\left(\frac{2p^2}{B}\frac{l}{p} + \frac{p^2}{B}\right)\frac{mn}{p^2} = \frac{2\sqrt{3}lmn}{B\sqrt{M}} + \frac{mn}{B}.$$

# B I/O Lower Bound for a Chain of Matrix Multiplications

**Problem:** Given $n$ matrices $\mathbf{A}_i(d_i \times d_{i+1})$, $i = 1, \ldots, n$, compute their product $\mathbf{C} = \mathbf{A}_1\mathbf{A}_2 \cdots \mathbf{A}_n$. The available memory can hold $M$ scalar numbers. Suppose any matrix involved in the computation of $\mathbf{C}$ has size $\gg M$. Let $N$ be the number of scalar multiplications performed in order to compute $\mathbf{C}$. All input matrices and the result matrix reside on disk. Each disk block can store $B$ numbers. Give an optimal schedule that minimizes the amount of I/O in terms of disk blocks read/written.

**Solution:** We first bound the number of scalar multiplications that can happen during $M$ elements of I/Os. During $M$ elements of I/Os, the number of distinct elements that appear in memory must be $\leq 2M$. Define an *active* matrix multiplication to be $\mathbf{Z} = \mathbf{X}\mathbf{Y}$, where $\mathbf{X}$, $\mathbf{Y}$, and $\mathbf{Z}$ can be input matrices or intermediate results, and some elements of $\mathbf{X}$ and $\mathbf{Y}$ are in memory and produce some elements of $\mathbf{Z}$. There could be multiple active matrix multiplications in progress. Suppose $m_1, \ldots, m_k$ memory resource is allocated to each active matrix multiplication, so that $m_1 + \cdots + m_k \leq 2M$. According to the proof in Appendix A, at most

$$\left(\frac{2m_1}{3}\right)^{\frac{3}{2}} + \cdots + \left(\frac{2m_k}{3}\right)^{\frac{3}{2}} \leq \left(\frac{2m_1}{3} + \cdots + \frac{2m_k}{3}\right)^{\frac{3}{2}} = \left(\frac{4M}{3}\right)^{\frac{3}{2}}$$

multiplications can happen during $M$ number of I/Os.

It follows that the minimum number of element I/Os for computing $\mathbf{C}$ is

$$\frac{N}{\left(\frac{4M}{3}\right)^{\frac{3}{2}}} M = \Theta\left(\frac{N}{\sqrt{M}}\right).$$

Therefore, the I/O lower bound in terms of blocks is $\Theta\left(\frac{N}{B\sqrt{M}}\right)$.

The above reasoning also indicates that the lower bound is achieved by doing one active matrix multiplication at a time, and by applying the optimal schedule in Appendix A to each matrix multiplication. As an example, suppose we want to compute $\mathbf{C} = \mathbf{A}_1\mathbf{A}_2\mathbf{A}_3$, and among all possible ways of parenthesizations, $\mathbf{A}_1(\mathbf{A}_2\mathbf{A}_3)$ takes the minimum number of scalar multiplications. The optimal I/O performance is attained by first using the schedule in Appendix A to compute $\mathbf{T} = \mathbf{A}_2\mathbf{A}_3$ and materialize $\mathbf{T}$. Following that, $\mathbf{C} = \mathbf{A}_1\mathbf{T}$ is similarly computed.